\documentclass[10pt,twoside,reqno]{amsart}
\usepackage{mathbbol,mathtools,slashed}

\usepackage{multirow}
\usepackage[bookmarksnumbered, plainpages]{hyperref}
\usepackage[rightcaption]{sidecap}
\usepackage{caption,xparse}
\usepackage{pgfplots}

\usepackage{float}

\usepackage[english]{babel}

\usepackage[latin1]{inputenc}
\usetikzlibrary{positioning,arrows,patterns}

\usepackage{graphicx}
\usepackage{epstopdf}
\usepackage{subfig}
\usepackage{color}
\usepackage{amsthm}
\usepackage{amsmath, amsfonts, amssymb}
\usepackage[figuresright]{rotating}
\begin{document}
   \title{Thermal propagator of the bosons and fermions fields}
   \author{{\small M. A. A. AHMED$^{1,2,\dag}$
    H. ZAINUDDIN$^{1,3,\ddag}$, N. M. Shah$^{1,3,\star}$.}}
    \thanks{\scriptsize
 emails:\\ $^\dag$ mohammed\_h7@$_{\text{yahoo.com}}^{\text{taiz.edu.ye}}$\\$^\ddag$ hisham@upm.edu.my.\\ $^\star$risya@upm.edu.my}
 \maketitle
\begin{center}
\address{$^{1}$ Laboratory of Computational Sciences and Mathematical Physics, Institute for Mathematical Research, Universiti Putra Malaysia, UPM Serdang 43000, Selangor, Malaysia\\
 $^{2}$Physics Department, Faculty of Science, Taiz University Al-Turba branch, Taiz, Yemen.\\
 $^{4}$ Department of Physics, Faculty of Science, Universiti Putra Malaysia, 43400 Serdang, Selangor, Malaysia.}
\end{center}
\begin{abstract} 
In a thermodynamic environment, thermal field theory (TFT) describes a large ensemble of interacting particles. This may appear to be the same as in classical statistical mechanics. Therefore, we study the scalar propagator and fermion propagator by considering an arbitrary parameter $\sigma$ that allows for a path integral description in real-time formalism (RTF). We constructed those propagators which allowed us to know how particles moved from one point to another point in momentum space as well as in mixed space, at finite temperature without a chemical potential.
\end{abstract}
\markboth{\rightline {\sl M. A. A. Ahmed et al. }}
        {\leftline{\sl  Thermal propagator of the bosons and fermions fields}}

\section{INTRODUCTION}
There are two formalisms within thermal field theories (TFT) for real time. So, the Real-Time Formalism (RTF) can be made fully consistent according to a matrix structure. This structure allows one to incorporate an arbitrary parameter for the family of paths for the general time path contour in the complex $t$ plane. We should point out that the two most common values for the $\sigma$ parameter that appears in the Feynman propagator rules are $(\sigma=0,1/2)$. Only for $\sigma=0$, it is called for the closed time path (CTP) formalism \cite{Schwinger11,Keldysh9}.For ThermoField Dynamics (TFD), it is associated to $\sigma=1/2$; this choice of $\sigma$ reproduces instead of the Feynman rules of an operatorial approach to quantum thermal field theory known as TFD \cite{Takahashi12}.
In this work, we present internal matrices for scalar boson and fermion fields in RTF without chemical potential \cite{Aurenche92}. In RTF, the matrix constructed connects the doubling of field degrees of freedom of the nonthermal vacuum to those of the thermal vacuum thermal vacuum, leading to the thermal propagator for the real massive Klein-Gordon theory as an expectation value in this thermal vacuum \cite{DAS2016}. Several works introduce the propagator system in the mixed space, showing an unexpected simple relation between any finite temperature $T\neq0$ graph and it's zero temperature $T=0$ counterpart through a multiplicative scalar operator that carried the entire temperature dependence \cite{das2006b}. This is done in RTF when ($\sigma=0$ and for it's conjugate $\sigma=1$).

The aim of the present manuscript is the computation of the thermal propagator components for the bosons and fermions in different spaces \cite{das2018,Lundberg2021}.

\section{GENERAL THERMAL PROPAGATOR OF PARTICLES IN RTF}

In thermal field theories, there are two commonly used formalisms in RTF,  are CTP and TFD formalisms. The path integral method is used in the CTP formalism, whereas TFD is based on an operatorial description of thermal quantum field theory. The doubling of fields, resulting to $(2\times2)$ matrix for the propagator \cite{DAS} will be written as
\begin{equation}\label{pr20}
  i G^{(\sigma,1/2)}(P;\beta)=\left(\begin{array}{ll}
G_{}^{(\sigma,11)}(P;\beta)&G_{}^{(\sigma,12)}(P;\beta)\\
G_{}^{(\sigma,21)}(P;\beta)&G_{}^{(\sigma,22)}(P;\beta)\end{array}\right).
\end{equation}
It is well known that if one wants to work with real-time propagators at finite temperature, whose elements correspond to the thermal propagators, for which the subscripts $(1,2)$ refer to the two real branches of the time contour. There are two types of propagator components. The $``12"$ and $``21"$ are mixing components, which are unphysical since one of the time arguments has an imaginary component. The only physical propagators are the $``11"$ and $``22"$ components. The TFT can be defined on a general time path in the complex $t$ plane, where $0\leq \sigma\leq1$ \cite{DAS2016}.
The Feynman propagator will be written here as $g(P)$ and its complex conjugate $g^{*}(P)$ is defined at zero temperature as
\begin{equation}\label{0pro}
g(P)=\frac{i(\slashed{P}+m)}{p_0^2-E^2+i\eta},
\end{equation}
where $E^2=\vec{p}^2+m^2$. Next, the matrix propagator at zero temperature is defined as
\begin{equation}\label{ch23}
 G(P)=\left(\begin{array}{cc}
 g(P)&0\\
0&g^{*}(P)\end{array}\right).
\end{equation}
Therefore we can use the properties
\begin{equation}\label{ch3pro1}
  g(P)+g^{*}(P)=\begin{cases}
 \displaystyle 2i\pi\delta(x_P^2)(\slashed{P}+m), & \displaystyle \mbox{for}\ \ \ \ \  \mbox{fermion} \\
 \displaystyle 2i\pi\delta(x_P^2), & \displaystyle \mbox{for}\ \ \ \ \ \mbox{boson},
                                    \end{cases}
\end{equation}
and
\begin{equation}\label{ch3pro2}
  g^{*}(P)-g(P)=\begin{cases}
 \displaystyle 2i(\slashed{P}+m)\mathcal{P}\frac{1}{x_P^2}, & \displaystyle \mbox{for}\ \ \ \ \  \mbox{fermion} \\
\displaystyle  2i\mathcal{P}\frac{1}{x_P^2}, & \displaystyle \mbox{for}\ \ \ \ \  \mbox{boson}.
                                     \end{cases}
\end{equation}
Note that $x_P^2:=(p_0)^2-E^2$, $\delta$ is the delta Dirac function and $\mathcal{P}$ is the cauchy principal value, $P=(p_0,\textbf{p})$ is in the momentum space, $P=(t,\textbf{p})$ is in the mixed space, and $\Theta(\pm p_0)$ denotes the step function. The matrix elements $G^{(\sigma,1/2)}_{}$ in the momentum space is given by:
\begin{equation}\label{ch26}
 \begin{array}{ll}
iG^{(\sigma,11)}_{}(P;\beta)=(\slashed{P}+m)\left[ \frac{i}{p_0^2-E^2+i\eta}+2\pi \delta(x_p^2)N_{(\xi)}(P;\beta)\right], \\
iG^{(\sigma,12)}_{}(P;\beta)=(\slashed{P}+m)\left(\Theta(-p_0)+N_{(\xi)}(P;\beta)\right)2\pi \delta(x_p^2)e^{\sigma\beta P},\\
iG^{(\sigma,21)}_{}(P;\beta)=(\slashed{P}+m)\left(\Theta(p_0)+N_{(\xi)}(P;\beta)\right)2\pi \delta(x_p^2)e^{-\sigma\beta P},\\
iG^{(\sigma,22)}_{}(P;\beta)=(\slashed{P}+m)\left[ \frac{-i}{p_0^2-E^2-i\eta}+2\pi \delta(x_p^2)N_{(\xi)}(P;\beta)\right],
\end{array}
 \end{equation}
 and $N_{\xi}(P;\beta)$ is the distribution of particles of the form
\begin{equation}\label{dis2}
N_{(\xi)}(P;\beta)=\frac{\Theta(p_0)}{e^{\beta(p_0)}-\xi}+\frac{\Theta(-p_0)}{e^{-\beta p_0}-\xi},
\end{equation}
where $\xi=1$ for bosons and $\xi=-1$ for fermions. There are several things to take note from the structures of the propagators' terms in (\ref{ch26}). Some of these terms are independent of temperature, present in $\beta$, while other terms are temperature-dependent $T\neq0$. Most importantly, in this section, $\sigma$ is an arbitrary parameter which appears in the exponential function in $G^{(\sigma,12)}_{}(P;\beta)$ and $G^{(\sigma,21)}_{}(P;\beta)$ for both fields (scalar and fermion). We show the relations between TFD and CTP, in propagators which depend on the values of $\sigma$.
Hence, the definition of the propagator in the mixed space, that is computed the elements of (\ref{ch26}) in this space to be
\begin{equation}\label{ch27}
G^{(\sigma,ij)}(t,\textbf{p};\beta)=\int_{-\infty}^{\infty}\frac{dp_0}{2\pi}\exp^{-ip_0t}G^{(\sigma,ij)}(p_0,\textbf{p};\beta),
\end{equation}
where $i,j\in\{1,2\}$ \cite{das2018}.

 \section{THERMAL PROPAGATOR OF SCALAR FIELD}

The free Lagrangian density for a scalar field without a chemical potential has been written in QFT as
\begin{equation}\label{ch28}
  \mathcal{L}(x)=\frac{1}{2}\left(\partial_\mu\phi(x)\right)^2-\frac{1}{2}m^2\phi^2(x).
\end{equation}
We use (\ref{0pro}) to write the free scalar propagator at zero temperature
$$\Delta(P)=\frac{i}{p_0^2-E^2+i\eta},$$
and the following propagator at finite temperature in (\ref{ch26}) for the scalar field
\begin{equation}\label{scalar}
\begin{array}{ll}
D^{(\sigma,11)}_{}(P;\beta)=\left[ -i\Delta(P)+2\pi \delta(p_0^2-E^2)N_{B}(P;\beta)\right], \\
D^{(\sigma,12)}_{}(P;\beta)=-2i\pi \delta(p_0^2-E^2)\left(\Theta(-p_0)+N_{B}(P;\beta)\right)e^{\sigma\beta P},\\
D^{(\sigma,21)}_{}(P;\beta)=-2i\pi \delta(p_0^2-E^2)\left(\Theta(p_0)+N_{B}(P;\beta)\right)e^{-\sigma\beta P},\\
D^{(\sigma,22)}_{}(P;\beta)=\left[ -i\Delta^*(P)+2\pi \delta(p_0^2-E^2)N_{B}(P;\beta)\right].
\end{array}
\end{equation}
The Boson-Einstein distribution can be written as hyperbolic functions by
$$\sinh^2 \theta(p_0;\beta)=\frac{1}{e^{\beta|p_0|}-1},$$
where
$$\cosh \theta(p_0;\beta)=\frac{e^{\frac{\beta p_0}{2}}}{\sqrt{e^{\beta|p_0|}-1}},  \ \ \ \ \ \text{and}\ \ \ \ \ \sinh \theta(p_0;\beta)=\frac{1}{\sqrt{e^{\beta|p_0|}-1}},$$
and
$$\sinh \theta(p_0;\beta)\cosh \theta(p_0;\beta)=N_B(p_0;\beta)f_B^{\left(\frac{1}{2}\right)}(p_0;\beta).$$
For a general path, one can make a specific choice of the arbitrary function, $f_B^{(\sigma)}(p_0;\beta)$ to considerably reduce the algebra. Setting for example $f_B^{(0)}(p_0;\beta)$ for CTP and $f_B^{\left(\frac{1}{2}\right)}(p_0;\beta)$ for TFD, one obtains,
\begin{equation}\label{fun0}
  f_B^{\sigma}(p_0;\beta)= \begin{cases}
  e^{\beta \frac{|p_0|}{2}},  &\sigma=\frac{1}{2} \ \ \text{for} \ \ \text{TFD}, \\
  \\
   1, & \sigma=0 \ \ \text{for} \ \ \text{CTP},
\end{cases}
\end{equation}
where $N_B(|p_{0}|;\beta)=\sinh^2\theta(p_0;\beta)=\frac{1}{e^{\beta| p_0|}-1}$ is the Boson-Einstein distribution function. We can note that in (\ref{scalar}) the system allow us to study the thermal effect and the particles interactions of the scalar field in both approaches of CTP and TFD by giving the values for $0\leq\sigma\leq1$ and the variable $(\beta)$ whose results agree with \cite{das2018,ahmed2021b}.
Furthermore, the propagator acquires a $2\times2$ matrix structure that allows one to study of the matrix elements in the mixed space by using (\ref{ch27}),
\begin{equation}\label{ch029}
 \begin{array}{ll}
D_{}^{(\sigma,11)}(t,\textbf{p};\beta)=\frac{1}{2E}\left[(\Theta(t)+N_B(E;\beta))e^{-itE}+(\Theta(-t)+N_B(E;\beta))e^{itE}\right], \\
D_{}^{(\sigma,12)}(t,\textbf{p};\beta)=\frac{1}{2E}\left[(\Theta(-E)+N_B(E;\beta))e^{-iE(t+i\beta\sigma)}+(N_B(E;\beta)+\Theta(E))e^{iE(t+i\beta\sigma)}\right], \\
D_{}^{(\sigma,21)}(t,\textbf{p};\beta)=\frac{1}{2E}\left[\Theta(E)+(N_B(E;\beta))e^{-iE(t-i\beta\sigma)}+(N_B(E;\beta)+\Theta(-E))e^{iE(t-i\beta\sigma)}\right], \\
D_{}^{(\sigma,22)}(t,\textbf{p};\beta)=\frac{1}{2E}\left[(\Theta(t)+N_B(E;\beta))e^{itE}+(\Theta(-t)+N_B(E;\beta))e^{-itE}\right], \\
\end{array}
 \end{equation}
which is in agreement with \cite{das2018,ahmed2021b} (for more details see appendix \ref{a}). To study the fermion field, we can follow similar steps used with the associated scalar field, in the momentum space, leading to a matrix factorization of the propagator in the mixed space.

 \section{THERMAL PROPAGATOR OF FERMION FIELD}

The free Lagrangian density for a fermion field without a chemical potential is
\begin{equation}\label{ch210}
  \mathcal{L}(x)=\bar{\psi}(x)(i\slashed{D}-m)\psi(x).
\end{equation}
The covariant derivative in the fundamental representation is $D_{\mu}\psi(x)=\partial_t\psi(x)-ie A_{\mu}\psi(x)$. The free-fermion propagator at zero temperature is given by
$$S(P)=\frac{i(\slashed{P}+m)}{p_0^2-E^2+i\eta},$$
where $\gamma$ is gamma matrix and $\slashed{P}=\gamma^{\mu} P_{\mu}=\gamma^0 p_0-\vec{\gamma}\cdot\textbf{p}$. We next use (\ref{dis2}) for fermion to get
$$\sin^2\theta(p_0;\beta)=N_F(p_0;\beta)=\frac{\Theta(p_0)}{e^{\beta p_0}+1}+\frac{\Theta(-p_0)}{e^{-\beta p_0}+1},$$
where $N_F(p_0;\beta)$ is the Fermi-Dirac distribution function. The goniometric functions are
$$\cos \theta(p_0;\beta)=\Theta(p_0)\sqrt{\frac{e^{\beta p_0}}{e^{\beta p_0}+1}}+\Theta(-p_0)\sqrt{\frac{1}{e^{\beta p_0}+1}},$$
and
$$\sin\theta(p_0;\beta)=\Theta(p_0)\sqrt{\frac{1}{e^{\beta p_0}+1}}+\Theta(-p_0)\sqrt{\frac{e^{\beta p_0}}{e^{\beta p_0}+1}}.$$
Now we can use (\ref{ch26}) to write the thermal elements of the propagator for fermion field as
\begin{equation}\label{fermion}
\begin{array}{l}
iS^{(\sigma,11)}_{}(P;\beta)=(\slashed{P}+m)\left[ \Delta(P)-2\pi \delta(p_0^2-E^2)N_{F}(P;\beta)\right], \\
iS^{(\sigma,12)}_{}(P;\beta)=2\pi (\slashed{P}+m)\delta(p_0^2-E^2)\left(\Theta(-p_0)-N_{F}(P;\beta)\right)e^{\sigma\beta P},\\
iS^{(\sigma,21)}_{}(P;\beta)=2\pi (\slashed{P}+m) \delta(p_0)^2-E^2)\left(\Theta(p_0)-N_{F}(P;\beta)\right)e^{-\sigma\beta},\\
iS^{(\sigma,22)}_{}(P;\beta)=(\slashed{P}+m+)\left[ -\Delta^*(P)-2\pi \delta(p_0^2-E^2)N_{F}(P;\beta)\right],
\end{array}
\end{equation}
Note that the previous study involved the hyperbolic functions for the scalar field, but here we use the goniometric functions to construct the fermion fields,
$$\sin \theta(p_0;\beta)\cos \theta(p_0;\beta)=N_F(p_0;\beta)f_F^{\sigma}(p_0;\beta).$$
The arbitrary function, $f_F^{(\sigma)}(p_0;\beta)$ reduce the algebra. We will clarify how the choice of approach affect on the function $f_F^{(0)}(p_0;\beta)$ for CTP and $f_F^{\left(\frac{1}{2}\right)}(p_0;\beta)$ for TFD, namely,
\begin{equation}\label{fun1}
 f_F^{\sigma}(p_0;\beta)= \begin{cases}
  e^{\beta \frac{p_0}{2}}\left(\left[\Theta^2(p_0)+\Theta^2(-p_0)\right]+2\cosh \left(\frac{\beta}{2}p_0\right)\Theta(p_0)\Theta(-p_0)\right),  &\text{for}\ \text{TFD}, \\
  \\
   1, & \text{for}\ \text{CTP}.
\end{cases}
\end{equation}
Here, we rewrite Fermi-Dirac distribution as $N_F(|p_0|;\beta)=\sin^2\theta(p_0;\beta)=\frac{1}{e^{\beta |p_0|}+1}$. In (\ref{fermion}) the matrix is useful to study the thermal effect and the particles interactions of the fermion field in RTF within this range $0\leq\sigma\leq1$ \cite{DAS,das2006a}. Now, we will write the elements of the fermion propagator in the mixed space. The propagator acquires a $2\times 2$ matrix structure in (\ref{fermion}); with the help of (\ref{ch27}) we obtain
\begin{equation}\label{ch0211}
 \begin{array}{ll}
S_{}^{(\sigma,11)}(t,\textbf{p};\beta)=\frac{1}{2E}\left[(\Theta(t)-N_F(E;\beta))A(E)e^{-itE}+(\Theta(-t)-N_F(E;\beta))B(E)e^{itE}\right], \\
S_{}^{(\sigma,12)}(t,\textbf{p};\beta)=-\frac{1}{2E}\left[B(E)\left(N_F(E;\beta)-1\right)e^{iE(t+2i\beta\sigma)}+A(E)N_F(E;\beta)e^{-iE(t+2i\beta\sigma)}\right], \\
S_{}^{(\sigma,21)}(t,\textbf{p};\beta)=-\frac{1}{2E}\left[B(E)N_F(E;\beta)e^{iE(t+2i\beta\sigma)}+A(E)\left(N_F(E;\beta)-1\right)e^{-iE(t+2i\beta\sigma)}\right], \\
S_{}^{(\sigma,22)}(t,\textbf{p};\beta)=\frac{1}{2E}\left[(\Theta(t)-N_F(E;\beta))B(E)e^{itE}+(\Theta(-t)-N_F(E;\beta))A(E)e^{-itE}\right], \\
\end{array}
 \end{equation}
 where,
$$A(E)=\gamma^0 E-\vec{\gamma}\cdot \vec{p}+m,$$
$$B(E)=-(\gamma^0 E+\vec{\gamma}\cdot \vec{p}-m).$$
When $\sigma=0$ the results in (\ref{ch0211}) agreed with CTP in \cite{das2006b} within chemical potential, but in our case, the result was for any path within RTF.

 \section{CONCLUSION}

As we know well that there are two approaches in RTF. Therefore, several works gathered together those formalisms through two popular choices for the contour parameter $\sigma$, $(\sigma=0$ and $1/2)$  for both CTP and TFD, respectively. The viewpoint suggested here is to study the thermal propagator of the fermion field and the thermal propagator of the scalar field. The results show that the real-time propagators for scalar and fermion fields consist of a sum of two parts, the non-thermal part and the thermal part, respectively. It is to be noted that the thermal part is different for each component of the propagator in both spaces for our work in the general path, while the thermal part is the same for those components in CTP formalism.
The thermal part of the propagator represents an on-shell contribution (because of the delta function). In fact, the intuitive meaning of the temperature dependent correction is quite clear. In a hot medium, there is a distribution of real particles. Hence, we can compute an integration over a real, continuous, energy $p_0$, since there is now a severe mathematical singularity owing to the square of the delta function. The temperature dependent part merely represents the possibility that a particle, in addition to having virtual exchanges, can also emit or absorb a real particle of the medium. While we have $\delta$- function and the particle distribution as well as the temporal transform factor $e^{-ip_0 t}$ in the mixed space, so the integration of those functions over $p_0$ will lead to the separation of the time and the spatial component of momentum.

 \section{ACKNOWLEDGMENTS}

HZ would like to acknowledge the Fundamental Research Grant Scheme (FRGS) under Ministry of Higher Education (MOHE) with project number FRGS/1/2019/STG02/UPM/02/3. NMS would like to acknowledge support from FRGS grant under Ministry of Higher Education (MOHE) with project number FRGS/1/2018/STG02/UPM/02/10.

\appendix
\section{}\label{a}
We introduce the integrations of some quantities:
$$\int_{-\infty}^{\infty}\frac{i}{p_0^2-E^2-i\eta}\frac{dp_0}{2\pi}e^{-ip_0t}$$
$$=\int_{-\infty}^{\infty}\frac{1}{2E}\left[\frac{i}{p_0+E+i\eta}+\frac{i}{p_0-E-i\eta}\right]\frac{dp_0}{2\pi}e^{-ip_0t}$$
 $$\int_{-\infty}^{\infty}\left[\frac{i}{p_0-E-i\eta}\right]\frac{dp_0}{2\pi}e^{-ip_0t}$$
 $$=e^{-itE}\int_{-\infty-E}^{\infty-E}\left[\frac{i}{\chi-i\eta}\right] \frac{d \chi}{2\pi} e^{-i\chi t}=\lim_{\eta \to 0} \Theta(-t)e^{-itE-i \eta}$$
 and
 $$\int_{-\infty}^{\infty} 2\pi \delta(p_0^2-E^2)N_B(|p_0|;\beta)\frac{dp_0}{2\pi}e^{-ip_0t}$$
 $$=\frac{1}{2E}\int_{-\infty}^{\infty} (\delta(p_0-E)+\delta(p_0+E))N_B(|p_0|;\beta)dp_0 e^{-ip_0t}$$
 $$=\frac{1}{2E}\left[N_B(|E|;\beta)e^{itE}+N_B(|E|;\beta)e^{-itE}\right]$$
 and
 $$\int_{-\infty}^{\infty} \delta(p_0-E)f^\sigma(p_0;\beta)N(|p_0|;\beta)dp_0 e^{-ip_0t}=e^{-itE}f^\sigma(E;\beta)N(|E|;\beta).$$

\end{document}